\documentclass[aps,prb, superscriptaddress, preprint, amssym]{revtex4}
\usepackage{latexsym}
\usepackage{graphicx}
\usepackage{rotating}
\usepackage{longtable}
\usepackage[usenames]{color}
\usepackage[normalem]{ulem}

\input colordvi

\begin{document}

\title{Strain relaxation in small adsorbate islands: O on W(110)}

\author{T.~O.~Mente\c{s}}
\affiliation{Sincrotrone Trieste S.C.p.A., Basovizza-Trieste 34012, Italy}

\author{N.~Stoji\'{c}}
\affiliation{Scuola Internazionale Superiore di Studi Avanzati (SISSA), Via Beirut 2-4, 
Trieste, I-34014, Italy}
\affiliation{ INFM-CNR Democritos, Theory @ Elettra group,  Trieste, I-34014, Italy}

\author{N.~Binggeli}
\affiliation{Abdus Salam International Centre for Theoretical Physics, 
Strada Costiera 11, Trieste 34014, Italy}
\affiliation{ INFM-CNR Democritos, Theory @ Elettra group,  Trieste, I-34014, Italy}

\author{M.~A.~Ni\~{n}o}
\affiliation{Sincrotrone Trieste S.C.p.A., Basovizza-Trieste 34012, Italy}

\author{A.~Locatelli}
\affiliation{Sincrotrone Trieste S.C.p.A., Basovizza-Trieste 34012, Italy}

\author{L.~Aballe}
\affiliation{CELLS-ALBA Synchrotron Light Facility, C3 Campus Universitat 
Aut\'{o}noma de Barcelona, 08193 Bellaterra, Barcelona, Spain}

\author{M.~Kiskinova}
\affiliation{Sincrotrone Trieste S.C.p.A., Basovizza-Trieste 34012, Italy}

\author{E.~Bauer}
\affiliation{Department of Physics, Arizona State University, Tempe, Arizona 85287-1504}

\date{\today}

\begin{abstract}
The stress-induced lattice changes in a p($1\times2$) ordered
oxygen layer on W(110) are measured by low-energy electron diffraction.
We have observed that small oxygen islands show a mismatch
with the underlying lattice. Our results indicate that 
along $[1\bar{1}0]$ the average mismatch scales inversely
with the island size as $1/L$ for all oxygen coverages up to 0.5 ML, 
while along $[001]$ it is
significant only for the smallest oxygen islands and scales as
a higher power of the inverse island size. The behaviour along
$[1\bar{1}0]$ is described by a one-dimensional finite-size
Frenkel-Kontorova model. Using this model, together with calculated
force constants, we make a quantitative
estimate for the change of surface-stress upon oxygen adsorption.
The result is consistent with our ab-initio calculations, which
give a relative compressive stress of -4.72~N/m along $[1\bar{1}0]$ and
a minute relative tensile stress of 0.15~N/m along $[001]$.
The scaling along $[001]$ is qualitatively explained as an effect
induced by the lattice relaxation in the $[1\bar{1}0]$ direction.

\end{abstract}


\maketitle

\section{Introduction}

Surfaces of solids present interesting phenomena owing to their reduced dimensionality and
nontrivial symmetry, having the semi-infinite bulk on one side and 
vacuum on the other. From a simplified point of view,
the charge freed up by the absence of the vacuum-side bonds has to be redistributed,
resulting in major changes in the elastic,\cite{ibach97} magnetic,\cite{stojic06} 
and other thermodynamic\cite{surface_melting} properties of the surface as compared to the bulk. 
Often this results in structural modulations 
in order to minimize the energy and reduce the stress on the surface.
On the other hand, adsorbates may induce similar effects, restructuring the surface 
at atomic or mesoscopic length scales.\cite{ibach97,note_stripe}
Our study is based on such observations,
and aims at a better understanding of the elastic properties of 
adsorbate-covered crystal surfaces.

Determination of surface stress, which is central to understanding 
self-organization processes on crystal surfaces, has proven to be a challenge. 
Especially, the few studies that compare experiment and theory
do not provide a systematic level of agreement. Two such comparisons
are on the change of surface stress upon oxygen adsorption on 
Pt(111)\cite{ibach97,feibelman97} and Cu(100).\cite{harrison06,prevot06}
In the first case, the calculation result is somewhat lower than the experimental one, 
whereas in the case of O/Cu(100) the theory exceeds the experimental results by a factor
as large as 3. In a similar study, the calculation significantly underestimates the experiment 
on O/Ni(100).\cite{hong04}
The uncertainties are attributed to 
the sensitivity to the boundary conditions in the setup and the
macroscopic nature of the measurement in the crystal bending experiments,
by far the most popular technique regarding adsorbate induced surface stress, and 
the lack of convergence in the density functional theory (DFT) calculations. 
Therefore, additional means of measuring 
surface stress are very important.

To this end, one could consider the lattice relaxations upon stress release at a boundary.
Strain relaxation in finite-size objects is not surprising, as it is reasonable 
to expect that the boundaries (be it one-dimensional or two-dimensional) will
assume a configuration minimizing the forces. Indeed, pioneering experimental studies
regarding the change of the lattice constant of tiny three dimensional crystallites were
done in the early 1950s,\cite{berry52} and somewhat later the connection of this observation 
to surface stress was established.\cite{vermaak68} In these studies the contraction of 
small spherical metal crystals was explained through the radial forces applied by the
surface to the inner part of the particles, laying out the basis for understanding our 
experimental observations.

Regarding surfaces, there is a wide range of studies on elastic effects due
to a variety of defects such as single atoms, atomic steps, and three dimensional adsorbate
islands. Within the general framework, defects are modeled as local
forces that induce long-range elastic relaxations on the surface and in the bulk
(see Ref.~\onlinecite{muller04} and references therein). The propagation of the displacements
into the bulk has been demonstrated most notably by x-ray diffraction analysis.\cite{croset02}
However, the relevant interactions are often confined to the most superficial layers, 
as in the case of O adsorbed on Pt(110)-(1$\times$2).\cite{helveg07}

Following this argument, to a first approximation, an adsorbate island 
can be considered as a two-dimensional crystallite
with a one-dimensional boundary. In the case of pseudomorphism, 
the adsorbate lattice locks on to that of the underlying crystal.
However, for small monolayer islands, the stressed layer might be able
to -at least partially- relax through its boundaries, and we can expect a
mismatch between the adsorbate and the substrate. 
There is already experimental work pointing to such relaxations both on metals
and on semiconductors, using real space imaging and 
diffraction techniques.\cite{helveg07,massies93,muller95,fassbender95} 
On the theoretical side, recent simulations show
deviations from pseudomorphic positions for small
two-dimensional islands.\cite{lysenko02, stepanyuk01} 

In this paper, we demonstrate that quantitative information on surface stress 
can be obtained by measuring such deviations from pseudomorphism as a function 
of the size of the monolayer adsorbate islands. 
We observe that small islands of oxygen on W(110) 
behave similarly to the cases mentioned above 
showing a mismatch to the underlying lattice, $a_{0}$, depending on their size
and the crystallographic direction within the surface plane.
Taking advantage of the distinct low-energy electron diffraction (LEED)
spots corresponding to the p($1\times2$) order, we follow the changes in the
average lattice spacing as a function of the average domain size obtained from the spot widths.
We find a mismatch scaling as the inverse island size, $1/L$, in the $[1\bar{1}0]$ direction,
while it is dominated by quadratic and higher order terms, $O((a_{0}/L)^{2})$, along $[001]$.
The lattice mismatch is explained as a strain relaxation, and it is described in terms of 
a Frenkel-Kontorova model. From this model, using calculated force constants,
we extract the surface stress value along $[1\bar{1}0]$.

Furthermore, we have performed density-functional theory (DFT) calculations in order to understand 
the nature of the oxygen induced surface stress change, obtained
by modeling the observed strain relaxations in our measurements.
While the early first-principles surface-stress calculations\cite{Nee87,NeeGod90} were mostly oriented
towards the understanding of the microscopic origin of stress, 
some of the more recent ones\cite{feibelman97,hong04,harrison06} are being directly compared
to the experimental measurements of the surface stress change, with varying success. 
Our {\em ab-initio} result agrees well with 
the stress value estimated from our model of the measured lattice relaxations in the 
$[1\bar{1}0]$ direction. In addition, the qualitatively different scaling of the lattice mismatch 
along $[001]$ can be understood from our calculation results, which show a lack of a significant 
stress change in this direction and a strain-stress coupling between the two orthogonal directions. 
The latter is expected to give rise to a $(1/L)^{2}$-like scaling along $[001]$, induced by the strain
relaxation in $[1\bar{1}0]$.

The paper is organized as follows:
section II describes the experimental setup and reports the LEED results 
on oxygen adsorption on W(110), section III presents 
the model of lattice relaxation as a balls-and-springs chain, and in
section IV, we describe the {\em ab-initio} calculations. In section V,
we present and compare results of our model and those of our DFT calculations
along with a discussion involving the available experimental and theoretical work on W(110).

\section{Experiment}

\subsection{Setup}

The micro-spot low energy electron diffraction ($\mu$-LEED) measurements were performed
with the SPELEEM microscope in operation at Elettra, Italy.\cite{locatelli06}
The instrument combines a variety of techniques including $\mu$-LEED and
low energy electron microscopy (LEEM). The specimen is illuminated with
an electron beam generated by a LaB$_{6}$ gun (energy width of $\sim$ 0.6~eV), 
at a flux density of less than 10$^{-2}$ nA/m$^{2}$. In LEED operation mode,
the microscope images the diffraction pattern produced by the sample at the
back focal plane of the objective lens.\cite{bauer94} The probed area on the surface
is selected by inserting an aperture, which defines an incident electron beam
of 2~$\mu$m in diameter.

\begin{figure}[!]
\begin{center}
\includegraphics[width=7cm]{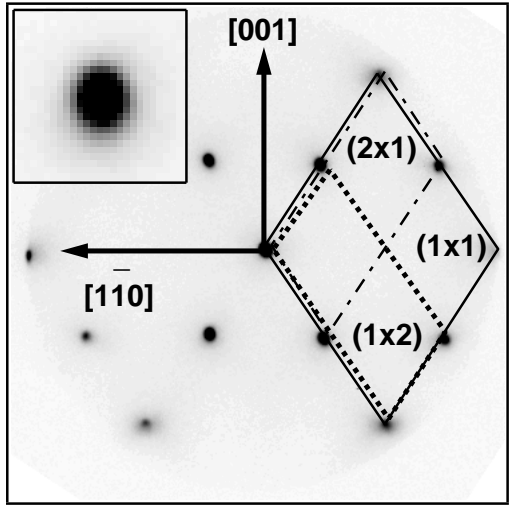}
\caption{The p($1\times2$) LEED pattern corresponding to 0.5 ML oxygen on W(110). 
The crystallographic directions are marked on the figure. 
Note the presence of both ($1\times2$) and ($2\times1$) domains. A blow-up of the
(0,-1/2) spot is shown in the inset.}
\label{fig:figureleed1x2}
\end{center}
\end{figure}

Prior to the $\mu$-LEED measurements, a micro-region with a single mono-atomic
W step was chosen using the LEEM mode, in order to avoid 
broadening of the diffraction peaks due to step bunches.
The reciprocal space calibration and the measurement of the transfer width
of the instrument were performed in such a region of the clean unreconstructed 
($1\times1$) W surface. We determined an instrument transfer width of about 110 \AA\
at 30~eV electron energy.

The W(110) crystal was cleaned by annealing at 1000~C in $2\times10^{-6}$~mbar oxygen and
subsequent high temperature flashes in ultrahigh vacuum to remove oxygen. 
The base pressure of the experimental
chamber was $2.5 \times 10^{-10}$ mbar. During the high temperature flashes, the
pressure remained below $3 \times 10^{-9}$ mbar. The sample was checked
using LEEM and LEED in order to confirm the absence of tungsten carbides forming 
on the surface. 

In order to form the adsorbed oxygen phases, 
molecular oxygen was dosed using a precision leak valve at a partial pressure of 
$5 \times 10^{-9}$ mbar as measured by an ion gauge. During the dosing, the LEED pattern 
was acquired every 20 sec, with the sample kept at 450 K. 
An electron energy of 30 eV was used for the diffraction measurements, corresponding to 
the intensity maximum of the half-order spots of the p($1\times2$) structure. 
For the oxygen pressure used, the best p($1\times2$) order
was obtained upon exposure of about 4.5 Langmuirs, 
assigned to 0.5 ML oxygen coverage.
The resulting LEED pattern is displayed in
Fig.~\ref{fig:figureleed1x2}. 

\subsection{LEED Measurements}

Adsorption of oxygen on W(110) has been thoroughly investigated in the past.\cite{engel75}
It is well known that for coverages below 0.5 ML, oxygen can be found in islands of 
p(1$\times$2) order, which evolves into p(2$\times$2) and (1$\times$1) reconstructions 
with increasing oxygen coverage.\cite{johnson93}
In spite of the stress induced by the oxygen layer,
the effective attractive interactions between the oxygen atoms manage to stabilize ordered islands
even at very low oxygen coverages (note that effective interactions include attractive nearest-neighbour,
repulsive next-nearest-neighbour of similar magnitude, and attractive but weaker
third-nearest-neighbour interactions\cite{wu89}).

\begin{figure}[!]
\begin{center}
\includegraphics{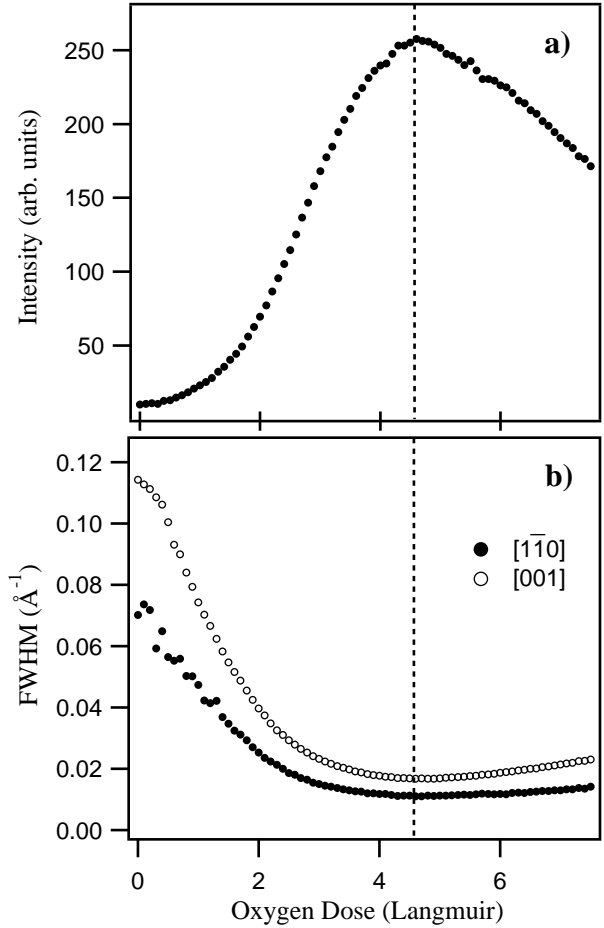}
\caption{The intensity and the full width at half maximum (FWHM) of the p($1\times2$) spots are
given as a function of the oxygen dose. The instrumental transfer width is accounted for the 
consideration of the FWHM values. The dashed line corresponds to 0.5 ML oxygen coverage.}
\label{fig:figureamp}
\end{center}
\end{figure}

Fig.~\ref{fig:figureamp} shows the evolution of the intensity and width of 
the p($1\times2$) half-order spots during oxygen uptake at 450 K. 
In agreement with the literature, 
even at the lowest coverages we can identify diffuse p($1\times2$) spots,
which grow in intensity and become sharper upon further exposure to oxygen. 
As seen in Fig.~\ref{fig:figureamp}, the p($1\times2$) intensity maximum at 4.5~L
corresponds to a minimum of the spot widths. Above this coverage, the 
p($1\times2$) domains break up due to additional oxygen atoms filling in the missing 
oxygen rows, which results in a slight broadening of the spots. 
Interestingly, the half-order spot width along $[1\bar{1}0]$ is always
smaller compared to that along $[001]$ roughly by a factor of $\sqrt{2}$, 
pointing to oxygen islands elongated in the $[1\bar{1}0]$ direction.
Beyond 0.5~ML, a weak p($2\times2$) order appears, 
developing very slowly due to the drop in the 
oxygen sticking coefficient with increasing coverage.\cite{engel75}

In the analysis of the half-order spot profiles, both Gaussian and Lorentzian fits
were used in order to extract the full-width at half maximum
(FWHM) values and the spot positions. 
A slowly varying background was subtracted as a low order polynomial. 
At high coverages, close to 0.5~ML, the peak shapes are predominantly Gaussian with a weak 
Lorentzian tail, whereas at the lowest
coverages both fits gave equally good results. 
The Lorentzian tail at higher coverages is consistent with the results of 
Wu {\it et al.} pointing to a bimodal island-size distribution.\cite{wu89}
Nevertheless, for simplicity we will consider only the mean value of
the size distribution extracted from the width of the dominant Gaussian contribution. 

\begin{figure*}[t]
\begin{center}
\includegraphics{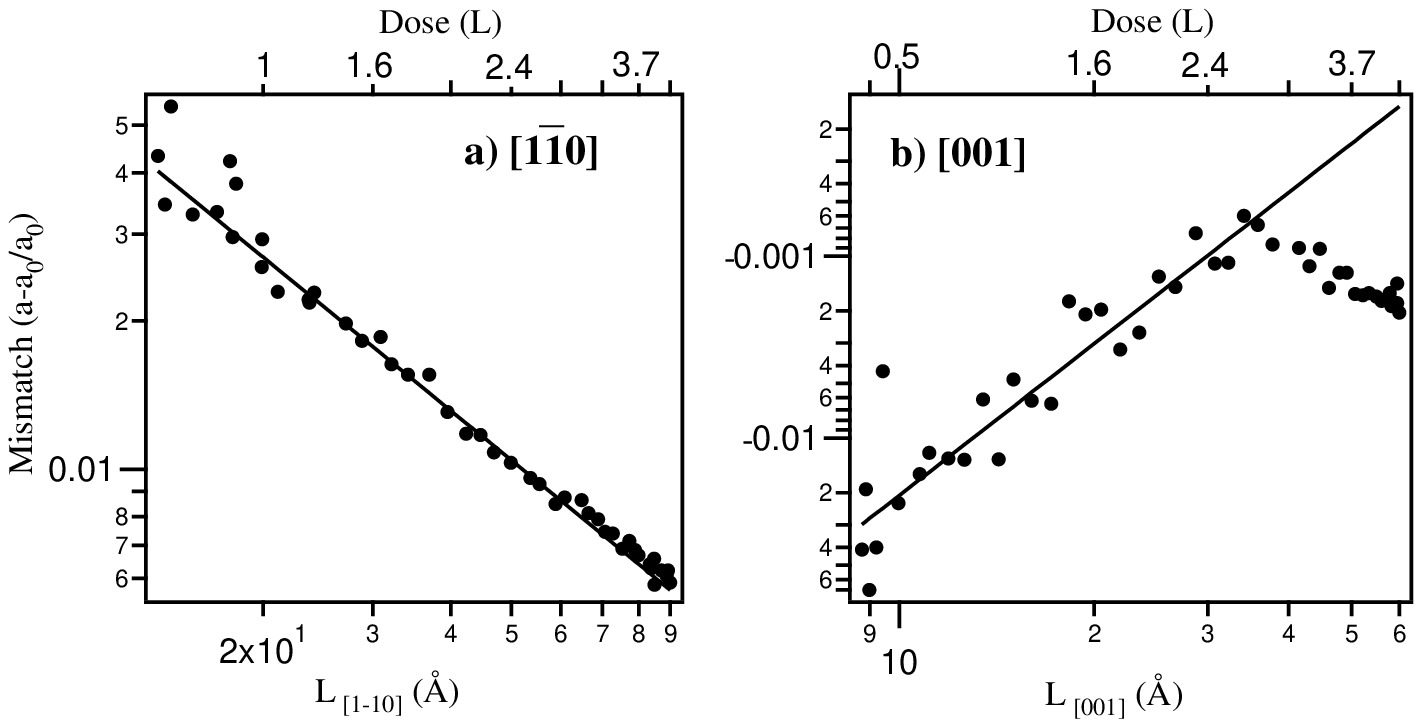}
\caption{The changes in the p($1\times2$) oxygen lattice spacing along $[1\bar{1}0]$ and 
$[001]$ are shown as a function of the average dimension of the ($1\times2$) domains. 
The filled circles are the data extracted from the LEED measurements, and the lines are
power law functions fitted to the data. Both axes are given in logarithmic scale.
The mismatch values are referenced to the W lattice. 
The domain sizes are found from the full width at half maximum of the diffraction spots. 
Note that the data are shown upto an oxygen coverage of 0.5 ML.}
\label{fig:figuremismatch}
\end{center}
\end{figure*}

Beyond the established behaviour of oxygen on W(110), we have observed a 
new effect manifested in the p($1\times2$) spot separations. Following the
evolution of the oxygen unit cell
as a function of oxygen coverage, we noticed deviations of the reciprocal
space vectors from those expected from the underlying tungsten lattice.
The mismatch of the oxygen p($1\times2$) unit cell to the tungsten lattice is displayed
in Fig.~\ref{fig:figuremismatch}. The fractional differences are
plotted against the average {\it domain size}, which is obtained from 
the FWHM values plotted in Fig.~\ref{fig:figureamp}b.
The variations are larger for low oxygen coverages also
corresponding to larger FWHM values of the half-order spots. We note that at low
coverages the oxygen lattice expands along $[1\bar{1}0]$ and shrinks
along $[001]$. The magnitude of the mismatch to the tungsten lattice 
is roughly an order of magnitude larger along $[1\bar{1}0]$, except at the lowest
oxygen coverages for which the p($1\times2$) LEED spots are barely visible.

The power law dependence of the mismatch on the domain size
is clear from the linear trend in the log-log plots in Fig.~\ref{fig:figuremismatch}.
The solid lines correspond to a fitting function of the kind:
\begin{equation}
\label{eq:eqn1}
 \frac{a_{meas}(L)-a_{0}}{a_{0}}\ =\ \frac{A}{L^{p}},
\end{equation}
where the left hand side corresponds to the measured average fractional mismatch of the oxygen
lattice with respect to the W(110) surface. $L$ is the linear size of the
oxygen island along the direction of interest, 
and $A$ and $p$ are the two parameters used to fit the data. 
The parameters resulting from the fit are displayed in Table~\ref{tab:table1}. 

\renewcommand{\baselinestretch}{1}
\begin{table}[ht]
\bigskip
\begin{center}
\begin{ruledtabular}
\begin{tabular}{ c c c }
  direction & \emph{A} (\AA$^{p}$)   & \emph{p}  \\
\hline
 $[1\bar{1}0]$                   & $0.56  \pm 0.05$     & $1.04 \pm 0.05$ \\
 $[001]$                         & $-0.78 \pm 0.21$     & $2.77 \pm 0.29$ \\
\end{tabular}
\end{ruledtabular}
\caption{Results of the power law fit to the data in Fig.~\ref{fig:figuremismatch}. 
The definition of the parameters $A$ and $p$ is given in Eq.~\ref{eq:eqn1}. }
\label{tab:table1}
\end{center}
\end{table}

As seen in Table~\ref{tab:table1}, along $[1\bar{1}0]$ the fractional
lattice mismatch scales inversely with the linear dimension $L$ ($p=1$). 
Along $[001]$, instead, the inverse scaling power
indicates a relaxation dominated by higher-order terms, i.e. $O((a_{0}/L)^{n})$ with $n \geq 2$.
Leaving the discussion to the following sections, here we limit ourselves to mention that
this points to a qualitative difference in the driving force of the strain relaxation
along the two directions. Apart from the scaling power, the particular value of the 
parameter $A$ carries more information specific to the adsorbate-substrate interactions,
especially along $[1\bar{1}0]$, as we will show in the next section.

In Fig.~\ref{fig:figuremismatch}b, we note that at $L_{[001]} \simeq 32$ \AA\ (corresponding
to an oxygen dose of 3~L and a coverage of 0.4~ML) the lattice along $[001]$
stops expanding, and starts shrinking. The likely origin of this abrupt change 
is an onset of ``island percolation'' or touching of the ordered islands.\cite{engel75} 
A kink is also present in Fig.~\ref{fig:figuremismatch}a at $L_{[1\bar{1}0]} \simeq 60$ \AA\
corresponding roughly to the same coverage. However, the effect is more evident in
Fig.~\ref{fig:figuremismatch}b due to the small magnitude of the fractional mismatch
along $[001]$.

\section{Modeling Strain Relaxation}

Assuming that lattice relaxation in finite-sized objects is limited to the
boundaries, one would expect that the average lattice spacing should converge
to a fixed value (to the bulk value for 3-dimensional crystals, or to the underlying lattice
for a 2-dimensional pseudomorphic film),
and that the difference from the bulk lattice should be inversely proportional to the object size. 
This statement is independent of the dimensionality, as the bulk divided by the boundary
always gives the linear dimensions of the system. The idea can be formulated through
a fractional mismatch:
\begin{equation}
\label{eq:eqn2}
 \epsilon\ \equiv\ \frac{a(L)-a_{0}}{a_{0}}\ \propto \frac{1}{L},
\end{equation}
where $L$ is the linear dimension along the direction under study, and $a_{0}$
is the lattice constant expected from an infinite object. One can safely assume that this 
simple scaling argument holds for larger objects. However, at the limit of a few atoms, 
this is not necessarily the case, as the interactions giving rise to stress may be
modified due to finite size effects.\cite{stepanyuk01}

Keeping this in mind, 
we note that for O/W(110) the average relaxation along $[1\bar{1}0]$ scales as $1/L$ 
all the way down to the smallest linear dimensions. Indeed, inverse scaling is
represented by the solid line in Fig.~\ref{fig:figuremismatch}a, which follows the
experimental points throughout the full range of domain sizes. 
The expansion of the oxygen lattice along $[1\bar{1}0]$ 
for the islands can be explained by the large compressive stress
in this direction from our {\em ab-initio} calculation for the p($1\times2$) 
O/W(110) surface, as we will show in Section V. 
The oxygen layer is compressed to match the W lattice along $[1\bar{1}0]$, 
and at the boundary of islands it relaxes the compression by expanding.

\begin{figure}[t]
\begin{center}
\includegraphics{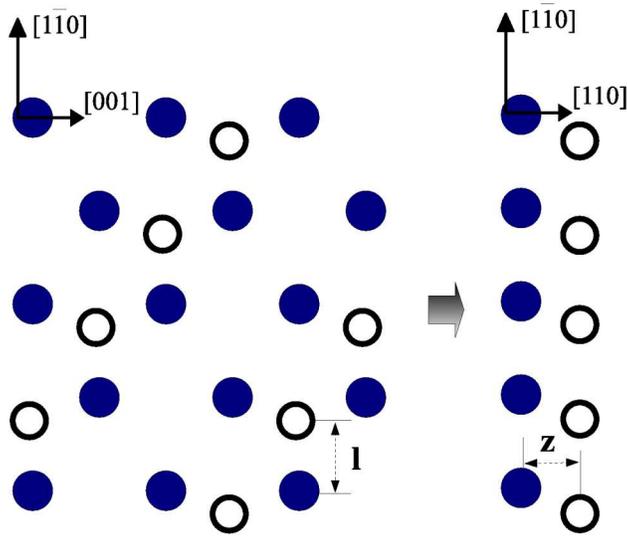}
\caption{The top and side views of  a p($1\times2$)-O covered area on W(110). 
Oxygen (tungsten) atoms are represented by the empty (solid) balls.  The two parameters, 
$z$ and $l$, describing the position of oxygen atoms are marked on the figure.}
\label{fig:figureFK_2d_to_1d}
\end{center}
\end{figure}

The almost perfect inverse scaling in the measurements encourages one to extract
quantitative information from the data using a simplified model. The information
specific to the system under study is coded in the proportionality factor in Eq.~\ref{eq:eqn2},
which we had termed as $A$ earlier in the experimental section. 
Having access to a single experimental coefficient, we will make a set of 
assumptions in order to reduce the number of parameters in the model problem:

\newcounter{Lcount}
\begin{list}{\roman{Lcount})}{\usecounter{Lcount}\setlength{\rightmargin}{\leftmargin}}
\item The displacements due to strain relaxation are assumed to be limited to 
the oxygen layer, with the tungsten atoms fixed.
\item $[1\bar{1}0]$ direction is treated independently of $[001]$.
\item The stress difference (across the island boundary) is assumed to be manifested 
as a point force at the island boundary within the surface plane.
\end{list}

The first assumption is a drastic one and is expected to introduce the largest
errors in the quantitative prediction of the model. Previous structural studies assumed
that the topmost tungsten layer  
preserves the atomic positions of the clean W(110) surface when covered with the 
(1$\times$2)-O layer,\cite{vanhove75,YnzDenPal00} pointing to a rigidity of the substrate
which provides the basis of our assumption.
However, the results of our DFT calculations, to be summarized in the next section, 
show that the W atomic positions are slightly modified upon oxygen adsorption 
both laterally and along the surface normal.

The second statement, on decoupling $[1\bar{1}0]$ from $[001]$, is valid due to the
small magnitude of oxygen-induced surface stress change along $[001]$.
This assumption leads to a reduction
(or projection) of the surface which can be understood from Fig.~\ref{fig:figureFK_2d_to_1d}. 
The ``W'' and ``O'' atoms on the side view of the surface  
can be regarded as effective particles corresponding to a line of atoms
along $[001]$. The force constants (describing the interactions at the harmonic limit)
along $[1\bar{1}0]$ between these effective atoms are defined by
infinitesimal movements of the corresponding line of atoms as a whole. 

\begin{figure}[t]
\begin{center}
\includegraphics{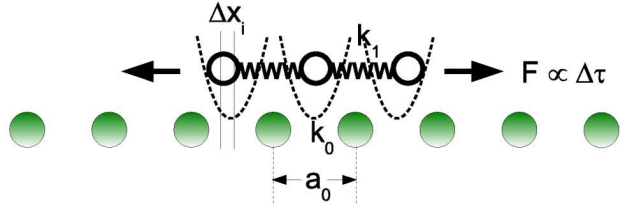}
\caption{The sketch of the side view of an oxygen island on the W surface. 
The oxygen atoms are denoted with empty circles, and the first layer tungsten atoms 
are shown as solid balls. The effect of the tungsten surface
on the adsorbate layer is represented by a periodic potential profile with
parabolic potential wells.}
\label{fig:figuresketchFK}
\end{center}
\end{figure}

In the third assumption, we replace the net force distribution at the island boundary
by a point force proportional to the surface stress difference between inside and outside the
islands.\cite{muller04,croset02} The resulting model for the adsorbate island is shown in
Fig.~\ref{fig:figuresketchFK}. The interactions between the oxygen atoms are sketched as
springs, whereas the effect of the tungsten substrate on the oxygen atoms is shown
as a periodic potential profile. Therefore, we have reduced the problem to that of 
a one-dimensional finite-size Frenkel-Kontorova chain.\cite{braun04}
Within such a model, only the nearest-neighbour interactions are considered.
We will make an additional assumption accounting for all interactions going beyond the
nearest-neighbours into an effective spring constant.

The relevant parameters in our model can be seen in Fig.~\ref{fig:figuresketchFK}. 
The substrate potential is defined by its period, $a_{0}$, and 
a force constant, $k_{0}$. The oxygen-oxygen bonds are modeled as springs,
with an equilibrium length, $a_{1}$, and a spring constant, $k_{1}$.
The relaxation of the oxygen-oxygen springs in the presence of the boundary 
is given by the distance of the $i^{th}$ oxygen atom to the $i^{th}$ adsorption site,
which we denote as $\Delta x_{i}$.
Assuming harmonic interactions, the total energy of a ($2N+1$)-atom adsorbate chain is
\begin{eqnarray}
\label{eq:eqn3}
          E_{tot} &=& 2 \sum_{i=1}^{N} \frac{1}{2}k_{0}(\Delta x_{i})^{2} +  \\
\nonumber        & & 2 \sum_{i=1}^{N} \frac{1}{2}k_{1}[a_{1}-(a_{0}+\Delta x_{i}-\Delta x_{i-1})]^{2}  ,
\end{eqnarray}
where we have assumed symmetry around the center with the center atom fixed ($\Delta x_{i=0} = 0$).
The positions in the ground state are found by setting the net force on each atom to zero 
($\partial E / \partial x_{i} = 0$). 

The fractional mismatch of the adsorbate chain to the substrate can be expressed as
\begin{equation}
\label{eq:eqn4}
 \epsilon = \frac{\Delta x_{N} - \Delta x_{-N}}{L} = \frac{2\Delta x_{N}}{L},
\end{equation}
where the length of the chain is $L=(2N+1)a_{0}$. The second equality is due to symmetry.
From a comparison of Eqs.~\ref{eq:eqn1} and \ref{eq:eqn4}, we see that $A = 2\Delta x_{N}$.
In words, the coefficient of the power law is directly related to the displacement of the
boundary atoms.

Although the statement of the problem looks simple, a general analytical solution is
surprisingly difficult to find. Instead, it is possible to approximate the solution 
for the mismatch 
\begin{equation}
\label{eq:eqn5}
 \epsilon = - \frac{2 F}{f(k_{0},k_{1})} \frac{1}{L} \ + \ O[(\frac{a_{0}}{L})^{2}],
\end{equation}
where $f(k_{0},k_{1})$ is a function which can be evaluated numerically (see Appendix).
It can be approximated as $f(k_{0},k_{1}) = k_{0} + ck_{1}$, where $c$ is a factor
of the order of unity depending on the ratio of the force constants ($c \approx 1$ for $k_0>>k_1$).
$F$ is the net force on the boundary atom corresponding 
to the change in the stress across the boundary. The higher order terms on 
the right contribute only for very small island sizes, $L < 4a_{0}$, above 
which they are negligible.
As expected for a one-dimensional system, the unit of stress is that of force. 
The connection to the actual stress on the two-dimensional surface along 
the direction of the chain can be established by
\begin{equation}
\label{eq:eqn7}
 F = a_{0\perp} \Delta\tau,
\end{equation}
where $a_{0\perp}$ is the unit length (defined per atom) along the perpendicular direction,
and $\Delta\tau$ is the change of the actual surface stress across the island boundary.
We note that, in this model, $F$ and $\Delta\tau$ do not depend on $L$, consistent
with an experimental $1/L$ scaling of $\epsilon$. Hence, $\Delta\tau$ corresponds to the
macroscopic surface stress induced by the chain in the asymptotic limit ($L\rightarrow\infty$).
Putting together Eqs.~\ref{eq:eqn5} and \ref{eq:eqn7}, we obtain the
measured coefficient as
\begin{equation}
\label{eq:modelresult}
 A = - \frac{2 a_{0\perp} \Delta\tau}{f(k_{0},k_{1})},
\end{equation}
which in essence states that the movement of the boundary atom is proportional to the net 
force on it, and is inversely proportional to the steepness of the potential well in which the
atom is sitting. Through this expression we have a means to relate the experimentally
measured power law coefficient, $A$, to the macroscopic surface stress.

In the next sections, we will combine the results of this model with
those from our {\em ab initio} calculations. In particular, we will
evaluate the microscopic force constants $k_{0}$ and $k_{1}$ from DFT
calculations. We will then compare the resulting stress from
Eq.~\ref{eq:modelresult} to the macroscopic surface stress obtained from the
first-principles calculations.

\section{{\em Ab-initio}  calculations}

We have performed DFT pseudopotential calculations in a plane-wave basis, 
using the PWSCF code.\cite{BarDalGir} The surface stress was calculated analytically 
on the basis of the expression derived by Nielsen and Martin,\cite{nieMar83}
based on the Hellmann-Feynman theorem, and following Ref.~\onlinecite{Nee87}.  
We utilized a symmetric slab with  
13 layers to simulate the W surface  and
15 layers for O/W, with 9 vacuum layers, both for relaxations and subsequent 
stress calculations. We used the local-density approximation (LDA)
in the Perdew-Zunger parametrization\cite{PerZun81} for exchange and correlation,
and employed Vanderbilt ultra-soft 
pseudopotentials\cite{Van90} generated from the 
$2s^2 2p^4$ atomic configuration of oxygen and  $5s^2 5p^6 5d^4 6s^2$ configuration 
of tungsten. The pseudopotential core-cutoff radii  for O were: $r_{s,p}=1.6$, $r_d = 1.4$~a.u. and 
for W: $r_{s,d}=2.2$, $r_d = 2.4$~a.u.
Our kinetic energy cutoff was 35~Ry for the wave functions and 350~Ry for the charge density. 
We used 326 k-points in the irreducible Brillouin zone. 
The forces were converged better than 1.7~mRy/\AA. 

As a prerequisite for the surface stress calculations, 
we relaxed the positions of the atoms in the W slab, keeping 
the central three layers fixed. Table~\ref{tab:W_relax} shows our results 
for the out-of-plane relaxations of the two topmost layers of a clean W slab together
with the available data from the literature.  
The theoretical results of Arnold {\em et al.}\cite{ArnHupBay97} are for a 9-layer slab, 
calculated with an all-electron method using the LDA.
Qian and H\" ubner\cite{QiaHub99}, within the same method, used a 5-layer slab and the
generalized gradient approximation for exchange and correlation. 
Similarly, Batirev {\em et al.}\cite{BatHerRen98} applied a modified LAPW method with LDA to a 5-layer W slab.
On the other hand, 
Ackland and Finnis\cite{AckFin86} utilized a semi-empirical model with parametrized energy functions.
We note a good agreement between our results, experiment and 
the previous {\em ab-initio} calculation on a slab of sufficient size,\cite{ArnHupBay97}
while the 5-layer slabs were too small 
and the semi-empirical model\cite{AckFin86}
gave rather poor result, in comparison with experiment. 

\renewcommand{\baselinestretch}{1}
\begin{table}[ht]
\bigskip
\begin{center}
\begin{ruledtabular}
\begin{tabular}{ c c c }
                                                             & $\Delta d_{12}$(\%)  & $\Delta d_{23}$(\%) \\
\hline
 our LDA result                                               & $-3.6$              & $0.2$  \\
 Arnold {\em et al.}\footnote{Ref.~\onlinecite{ArnHupBay97}} & $-3.6$              & $0.2$  \\
 Qian and H\" ubner\footnote{Ref.~\onlinecite{QiaHub99}}     & $-4.1$              & $-0.4$ \\
 Batirev {\em at al.}\footnote{Ref.~\onlinecite{BatHerRen98}}& $-0.8$              & $0.3$ \\
 Ackland and Finnis\footnote{Ref.~\onlinecite{AckFin86}}     & $-1.2$              & $0$    \\
 experiment$^a$                                              & $-3.1$              & $0$     \\
\end{tabular}
\end{ruledtabular}
\caption{Tungsten relaxation for two topmost layers given in percentage of the bulk lattice constant, 
as obtained in various calculations and the experiment. }
\label{tab:W_relax}
\end{center}
\end{table}

\renewcommand{\baselinestretch}{1}
\begin{table}[ht]
\bigskip
\begin{center}
\begin{ruledtabular}
\begin{tabular}{ c c c c c }
                  & $a_{bulk}$(\AA)      & $l/a$    & $z_1/a$     & $z_2/a$ \\
\hline
 our result       & $3.14$              & $0.52$   & $0.38$     & $0.36$  \\
 Za{\l}uska{\em et al.}\footnote{Ref.~\onlinecite{ZalKruRom01}} & $3.00$  & $0.56$  & $0.43$ & $0.43$  \\
 experiment\footnote{Ref.~\onlinecite{YnzDenPal00}}             & $3.16$  & $0.53$  & $0.40$  & $0.40$\\
\end{tabular}
\end{ruledtabular}
\caption{ Position of the oxygen atom after relaxing the slab. First column shows the 
W bulk lattice constant.
$z_{1}$ and $z_{2}$ denote the vertical distance of the oxygen atom to the two unequivalent tungsten atoms
with one and two oxygen neighbours, respectively.}
\label{tab:O_relax}
\end{center}
\end{table}

Similarly, we relaxed the atomic positions in the tungsten slab with adsorbed 
oxygen in the  p($1\times2$)-reconstruction.
Figure~\ref{fig:figureFK_2d_to_1d} gives top and side views of the p($1\times2$)-O/W(110) surface
and defines the  parameters $l$ and $z$ describing the position of the oxygen atom. 
The results of the calculation corresponding to these positional 
parameters are given in Table~\ref{tab:O_relax}. 
There, the $l$ and $z$ displacements in the p($1\times2$) reconstruction
are shown for the relaxed slab in reference to the calculated 
W bulk lattice constant, which is given in the first column.
For comparison, the quantum-mechanical DFT calculation of Za{\l}uska-Kotur {\em et al.}
using the atomic-cluster approach with only one O atom is also included in the table.
As evidenced by the bulk lattice constant, the cluster method gives 
a considerably compressed structure.
On the other hand, there is excellent agreement between our calculation and the
experimental results based on LEED measurements,\cite{YnzDenPal00}
especially regarding the W lattice constant and the lateral displacement of the oxygen atom.
In terms of layer-spacings, the LEED I(V) analysis\cite{YnzDenPal00}
did not allow any out-of-plane movement of the substrate atoms upon oxygen adsorption, 
while our relaxed positions indicate a $\sim0.05$~\AA~ buckling of the outermost plane of
W atoms, due to the oxygen atoms.
This result is consistent with an original estimate
of up to 0.05~\AA~ for the out-of-plane displacement due to oxygen adsorption.\cite{BucWanLag74} 
Additionally, the O atom in our calculation is displaced in the $[001]$ direction by less than 0.05~\AA, 
which was also not taken into account in the LEED model of Ref~\onlinecite{YnzDenPal00}. We note
that this rigid displacement of the whole adsorbate layer along $[001]$ is allowed
as it does not violate any symmetry operation on the surface.

\section{Surface Stress and Lattice Relaxation}

In order to get an estimate of $\Delta\tau$ along $[1\bar10]$
from the strain relaxation measurements using Eq.~\ref{eq:modelresult}, 
we have evaluated the force constants from the {\em ab-initio} slab calculation 
described in the previous section
by moving atoms infitesimally and following the change in energy.
The parameter $k_{0}$, which corresponds to the onsite potential profile
felt by the oxygen atoms, was found to be $k_{0} = 8.6$ eV/\AA$^{2}$ by
sliding the oxygen layer rigidly over the surface.\cite{note_forceconstant}
Similarly, the oxygen-oxygen force constant was obtained to be $k_{1} \approx 0.6$ eV/\AA$^{2}$
by moving every other oxygen atom along $[1\bar10]$ and taking into account 
the additional energy increase due to the substrate potential
(the total energy increase is proportional to $k_0+2k_1$).
The small value of $k_{1}$ with respect to $k_{0}$ is expected, as the
weak oxygen-oxygen nearest-neighbour interactions are further reduced by the
next-nearest-neighbour interaction of opposite sign and similar magnitude.\cite{wu89}
Taking the unit length along $[001]$ as $a_{0\perp}=6.33$ \AA, and the measured coefficient
$A = 0.56$ \AA, we obtain the oxygen induced change in the surface stress along $[1\bar10]$
to be $\Delta\tau^{[1\bar{1}0]} = -6.5$~N/m from Eq.~\ref{eq:modelresult}. 

We present our surface-stress results from {\em ab-initio} calculations on clean W(110) 
and on p($1\times2$)-O/W(110) in the first two columns in Table~\ref{tab:dft_stress}.
The difference of the two, which is the oxygen induced surface stress change, 
is listed in the third column. The numerical uncertainty of the calculated LDA
surface stress is estimated to be 0.36~N/m, on the basis of
convergence checks in which we increased the energy cutoffs and 
the number of k-points. To check further our results, we calculated the surface stress 
from the total energy difference for two strains of magnitude 1.5 \% and opposite signs
({\em i. e.} contraction and expansion of the lattice constant 
along [$1\bar10$]-direction), based on the energy expansion to the second order in the strain
(see Refs.~\onlinecite{BarKle65} and \onlinecite{WuHouZhu07}). By this method, 
we obtained a relative surface stress in the $[1\bar10]$ direction 
$\Delta\tau=-4.68$~N/m, which agrees very well with our
analytical result.

\renewcommand{\baselinestretch}{1}
\begin{table}[ht]
\bigskip
\begin{center}
\begin{ruledtabular}
\begin{tabular}{ c c c c c c }
  direction & $\tau_{W(110)}^{DFT}$   & $\tau_{p(1\times2)-O/W(110)}^{DFT}$  & $\Delta\tau^{DFT}$ & 
  $\Delta\tau^{LEED}$ &
  $\Delta\tau$\footnote{Ref.~\onlinecite{SanEndKir99} } \\
\hline
 $[1\bar{1}0]$   & $3.58$         & $-1.14$       &  $-4.72$     & $-6.5$     & $-1.1$  \\
  $[001]$        & $5.26$         & $5.41$        &  $ 0.15$     &  -    & $-0.1$  \\
\end{tabular}
\end{ruledtabular}
\caption{The results of our surface stress calculations and 
of the relative stress between O/W(110) and W(110) surfaces. 
For comparison we give in the last two columns the surface stress change obtained from
our LEED analysis and from a previous experiment using the crystal-bending method.\cite{SanEndKir99}
The values are all given in N/m. A positive sign corresponds to tensile stress. }
\label{tab:dft_stress}
\end{center}
\end{table}

The calculated surface stresses along $[1\bar{1}0]$ and $[001]$ of the 
clean W(110) surface are both tensile. This is in qualitative agreement 
with the previous theoretical estimates.\cite{BatHerRen98,AckFin86} 
We note, however, that our surface stress values are larger than the
average surface stress of 2.7~N/m obtained from a previous LDA
calculation, in Ref.~\onlinecite{BatHerRen98}, for a 5-layer slab. Aside from having the
correct atomic structure, stress values do converge much slower with the
number of k-points than energies in the DFT calculations, and the
long-ranged elastic relaxations bring forth the necessity for larger
slabs, which is the motivation behind the rather thick slabs used in our
calculations. The semi-empirical potential calculations\cite{AckFin86} yield
instead a surface stress of 2.4~N/m along $[1\bar{1}0]$ and 0.3~N/m along
$[001]$, which differ significantly from the present {\em ab initio}
results. Apart from the method, the difference also derives from the
poor structural optimization obtained in the empirical potential
approach. To the best of our knowledge, no experimental data are available for 
the clean W(110) surface stresses. However, experimental values do
exist for a related quantity which contributes to 
the surface stresses: the surface energy.\cite{Nee87} 
Our calculated surface energy of 3.5~J/m$^{2}$ is within
the range of the available experimental data, 2.8 - 3.7~J/m$^2$.\cite{RodBozFer93}.

As for the oxygen induced change in the surface stress,
the crystal-bending data\cite{SanEndKir99} have the same sign but differ considerably
in magnitude from our result. As seen in
Table~\ref{tab:dft_stress}, both our calculation and the crystal-bending
measurement show that the relative stress is compressive along $[1\bar{1}0]$, while
along $[001]$ the surface stress is affected negligibly upon oxygen adsorption.
On the other hand, the surface
stress change that we obtained by modeling the average lattice relaxation as a function
of island size, $\Delta\tau^{[1\bar{1}0]} = -6.5$~N/m, agrees well with the value $-4.72$~N/m
from the calculation, especially considering the few parameters used in our model. 

The reason for the remaining difference between the quantitative estimate from the model chain and
the calculated stress is attributed mainly to the simplifications of the model itself. Most
importantly, we have considered only the relaxations within the oxygen layer.
As hinted by the modified tungsten atomic positions in the presence of oxygen,
the relaxations in the adsorbate layer should continue into the tungsten surface, albeit
with reduced magnitude. Thus, a more accurate model should take into account
layer-resolved stresses, with the lattice in each layer (namely oxygen and the first layer 
tungsten) having a different power law behaviour (both with $p=1$ but each with a different 
coefficient, $A$). This, although possible, would require a more precise measurement and
a detailed analysis of the LEED spot profiles.

A quantitative comparison between the {\em ab-initio} calculations and the crystal-bending results 
from Table~\ref{tab:dft_stress} is less satisfactory, 
as the calculation is a factor four above the experimental value along $[1\bar{1}0]$.
This, although discouraging, illustrates
the problems in the surface stress studies.
One problem noted in the mentioned experimental study is the adsorption
of oxygen also on the back side of their crystal, which should result in a
reduction of the measured stress by about 25 \%.\cite{SanEndKir99} In addition,
as suggested in the introduction, part of the reason for the difference between
experiment and theory is to be sought in the macroscopic nature of the crystal-bending
measurements, since the presence of domain boundaries relieves
the stress on the adsorbate covered surface. 
Indeed, our LEED analysis shows that even at 0.5 ML oxygen coverage, which corresponds
to the best p($1\times2$) order, the oxygen lattice is relaxed along $[1\bar{1}0]$ 
by as much as 0.5 \%. At room temperature, this relaxation should be more
pronounced due to the smaller domain sizes. As a consequence, the surface stress in the adsorbed
face of the crystal is lower than the value estimated from a calculation that
considers a surface uniformly covered and free of defects. 
The surface quality of the crystal used in the experiments 
can also influence the results, as the surface stress would be 
reduced by the presence of defects such as steps or step-bunches. 
This problem was avoided in our $\mu$-LEED measurements
by choosing a region free of atomic steps. 

Regarding the strain relaxation processes, 
we have demonstrated that the behaviour along $[1\bar{1}0]$ is consistent with
what is expected from a harmonic chain sitting on a periodic potential.
However, along $[001]$ the situation presents a qualitative difference as the power 
(with which the relaxation scales inversely as a function of island size)
is much larger than 1 throughout the full range of oxygen coverages. 
Additionally, the mismatch along $[001]$
is much smaller than that along $[1\bar{1}0]$ except for the smallest
island sizes (see Fig.~\ref{fig:figuremismatch}).
As we have noted earlier, these observations 
suggest that the lattice relaxation along $[001]$ is a consequence of the $[1\bar{1}0]$ 
relaxation, yielding a higher-order effect in $a_0/L$.
Indeed, as seen in Table.~\ref{tab:dft_stress}, our calculations
predict no noticeable difference
between the clean and oxygen-covered W(110) surface stress in the $[001]$ direction. 
We suggest that the lattice change along $[001]$ is
induced by the strain-relaxations taking place in the $[1\bar{1}0]$ direction
for small-sized islands.
We find support for this explanation
when we extend the calculation to a slab that is stretched slightly along $[1\bar{1}0]$
in order to model the strain relaxation. In particular, we have performed
a calculation for a slab stretched uniaxially by 1.5 \% along $[1\bar{1}0]$.
In this case, the difference in the surface stress with and without the oxygen 
on the surface was found to be $\Delta\tau^{[1\bar{1}0]} = -3.98$~N/m and
$\Delta\tau^{[001]} = +0.68$~N/m.\cite{stretching_note} 
As expected, along $[1\bar{1}0]$ the relative stress decreases as the compressed lattice
relaxes upon stretching. On the other hand, along $[001]$ a tensile stress appears as 
the lattice is stretched in the orthogonal direction. This strain-stress coupling
between the two directions is therefore believed to be 
the driving force of the lattice changes along $[001]$.

A similar analysis as in Eq.~\ref{eq:eqn2} can give us more insight on the power law
behaviour observed along $[001]$. If the driving force is the relaxation
along $[1\bar{1}0]$, the regions in which the contraction in $[001]$ occurs must be limited
to what can be described as the ``corners'' of an island. It follows that there is an
inverse relationship of the relaxation along $[001]$ with the island area,
\begin{equation}
\label{eq:eqn9}
 \epsilon_{[001]}\ \propto \ \frac{1}{L^{2}}.
\end{equation}

The tendency of the strain relaxation, $\epsilon$, to decay faster along $[001]$ compared 
to $[1\bar{1}0]$ supports this argument. However,
the difference between the measured inverse power $p=2.76$ 
(see Table~\ref{tab:table1}) and the estimated inverse square relationship 
along $[001]$ is statistically significant. One issue that we have ignored in
the discussion of the spot profiles and relaxations is related to the island
size and shape distribution. In general the boundaries of oxygen islands
do not run simply along $[1\bar{1}0]$ and $[001]$. 
STM studies of low oxygen coverages show a variety of 
island shapes possibly elongated in the $\langle 1\bar{1}1 \rangle$ directions.\cite{johnson93}
However, from our two-dimensional LEED spot profiles, we were able to determine
only a mean elongation along $[1\bar{1}0]$.
In addition to the shape, the details of the island size distribution depend 
on the oxygen coverage.\cite{wu89} It is plausible that a coverage dependent change 
in the asymmetry of this distribution would modify the power law of
the apparent (or mean) strain relaxation.

\section{Conclusion}

We have measured the lattice relaxations within oxygen islands on W(110) by 
low-energy electron diffraction.
The half-order diffraction spots of the p($1\times2$) order within the oxygen
covered regions allowed the observation of the mismatch to the substrate lattice
as a function of the average island size.
Along $[1\bar{1}0]$, the mean lattice mismatch was shown to scale as
the inverse island size, $1/L$, while along $[001]$ the scaling involved
a higher power of the inverse island size.
We analysed the inverse-scaling behaviour along $[1\bar{1}0]$ through a
finite-size balls-and-springs chain in order to make an
estimate of the oxygen-induced surface stress change.
The result is in fair agreement with our density-functional theory calculation.
In addition, the calculations gave a small tensile stress along $[001]$,
which increased upon relaxing the compressive strain along $[1\bar{1}0]$.
This provides support to our explanation that the
qualitatively different behaviour in the $[001]$ direction is an effect
induced by the lattice relaxation along $[1\bar{1}0]$.

We believe that, with a more sophisticated analysis
of the integral order LEED spot profiles, the semi-quantitative
strain-relaxation model described in this paper
can be generalized to study the surface stress in a wide range of
adsorbate systems and ultra-thin films.


\appendix*
\section{}
The derivation of Eq.~\ref{eq:eqn5} follows from setting the net force on
each atom to zero in the Frenkel-Kontorova chain. Setting 
$\partial E / \partial x_{i} = 0$ (for $i=0,..,N$), we have,
\begin{eqnarray}
\label{eq:eqnAppendix1}
             (k_0 + 2k_1) \Delta x_i = k_1 (\Delta x_{i-1} + \Delta x_{i+1}) \ \ \ for\ &i&<N  \\
\nonumber F + (k_0 + k_1) \Delta x_N = k_1 \Delta x_{N-1} \ \ \ \ \ \ \ \ \ \ \ \ \ \ \ for\ &i&=N,
\end{eqnarray}
where $F$ is the net force due to the stress change on the boundary atom ($i=N$), 
before the relaxations, $\Delta x_i$, balance it. To find the displacements, we have to
solve simultaneously these $N+1$ equations. Instead of tackling the difficult
task of finding a general analytic solution, we can make the assumption that
only a few atoms from the boundary are displaced significantly. The result has the form
\begin{equation}
\label{eq:eqnAppendix2}
  \Delta x_N = - \frac{F}{k_0 + k_1 - \frac{k_1^2}{k_0+2k_1-\frac{k_1^2}{k_0+2k_1-...}}}
             = - \frac{F}{f(k_{0},k_{1})} ,
\end{equation}
where the last step defines the function $f(k_{0},k_{1}) \approx k_0 + c k_1$.
Independently, we solve the problem numerically by moving the atoms recursively until
all the forces vanish. A comparison of Eq.~\ref{eq:eqnAppendix2} with the numerical
solution gives a good agreement when we allow displacement of upto four atoms
from the boundary. The average mismatch in Eq.~\ref{eq:eqn5} follows directly from
Eqs.~\ref{eq:eqn4} and \ref{eq:eqnAppendix2}. \\

{\bf Note added:} During the refereeing process we became aware of an
article\cite{HarWooRob08} reporting {\em ab initio} calculations of surface
stress in a few adsorbate systems, including O/W(110). There is a good
agreement of their surface-stress values with ours.


\begin{thebibliography}{43}
\bibitem[{not({\natexlab{b}})}]{note_forceconstant}
\bibinfo{note}{We note that for the calculation of the force constants, we
  increased the kinetic energy cutoff to 50 Ry and the charge density cutoff to
  500 Ry.}
\bibitem[{str()}]{stretching_note}
\bibinfo{note}{Note that, for the stretched slab the calculated stress value
  has a large contribution from the bulk of the slab, which cancels out taking
  the difference between the two slabs with and without oxygen on the surface.}
\end{thebibliography}

\end{document}